\documentclass[12pt]{iopart}
\usepackage{iopams}  

\usepackage{amsmath,amsfonts,graphics,graphicx,epsfig,times,bbm}
\usepackage{amsfonts,graphics,times,bbm,amssymb}

\newtheorem{prop}{Proposition}\def\PRO{\begin{prop}}\def\ORP{\end{prop}}
\newtheorem{coro}{Corollary}\def\COR{\begin{coro}}\def\ROC{\end{coro}}
\newtheorem{theo}{Theorem}\def\TH{\begin{theo}}\def\HT{\end{theo}}
\newtheorem{defi}[prop]{Definition}\def\DE{\begin{defi}}\def\ED{\end{defi}}
\newtheorem{lemme}[prop]{Lemma}\def\LE{\begin{lemme}}\def\EL{\end{lemme}}

\def\ket#1{{|}#1\rangle}
\def\bra#1{\langle#1{|}}

\begin{document}

\title[Towards Minimal Resources of Measurement-based Quantum Computation]{Towards Minimal Resources of Measurement-based Quantum Computation}

\author{Simon Perdrix}
\address{PPS, CNRS - Universit\'e Paris 7}
\ead{simon.perdrix@pps.jussieu.fr}

\begin{abstract}
We improve the upper bound on the minimal resources required for measurement-based quantum computation \cite{N01,Leu04,Per05a}. Minimizing the resources required for this model is a key issue for experimental realization of a quantum computer based on projective measurements. This new upper bound allows also to reply in the negative to the open question presented in \cite{Per04a} about the existence of a trade-off between observable and ancillary qubits in measurement-based quantum computation.

\end{abstract}

\section{Introduction}
The discovery of new models of quantum computation (QC), such as the one-way quantum computer \cite{RBB02} and the projective
measurement-based model \cite{N01},  have opened up new experimental avenues 
toward the realisation of a quantum computer in laboratories. At the same time they have challenged 
the traditional view about the nature of quantum computation. 

Since the introduction of the quantum Turing machine by Deutsch \cite{Deu85a}, unitary transformations plays a central r\^ole in QC. However, it is known that the action of unitary gates can be simulated using
the process of quantum teleportation based on projective measurements-only \cite{N01}. Characterizing the minimal resources
that are sufficient for this model to be universal, is a key issue.

Resources of measurement-based quantum computations are composed of two ingredients: ($i$) a universal family of observables, which describes the measurements that can be performed ($ii$) the number of ancillary qubits used to simulate any unitary transformation.

Successive improvements of the upper bounds on
these minimal resources have been made by Leung and others \cite{FZ01, Leu04}. 
These bounds have been significantly reduced when the state transfer  (which is
a restricted form of teleportation) has been introduced: one two-qubit observable ($Z\otimes X$) and three one-qubit observables ($X$, $Z$ and $( X + Y)/ \sqrt 2$), associated with only one ancillary qubit, are sufficient for simulating any unitary-based QC \cite{Per05a}. Are these resources minimal ? In \cite{Per04a}, a sub-family of observables ($Z\otimes X$, $Z$, and $(X-Y)/\sqrt 2$) is proved to be universal,  
however two ancillary qubits are used to make this sub-family universal.

These two results lead to an open question : is there a trade-off between observables and ancillary qubits in measurement-based QC ? In this paper, we reply in the negative to this open question by proving that the sub-family $\{  Z\otimes X,Z,(X-Y)/\sqrt 2\}$ is universal using only one ancillary qubit, improving the upper bound on the minimal resources required for measurement-based QC.

\section{Measurement-based QC}

The simulation of a given unitary transformation $U$ by means of projective measurements can be decomposed into: 
\begin{itemize}
\item (\emph{Step of simulation}) First, $U$ is probabilistically simulated up to a Pauli operator, leading to $\sigma U$, where $\sigma$ is either identity or a Pauli operator $\sigma_x, \sigma_y$, or $\sigma_z$. 
\item (\emph{Correction}) Then, a corrective strategy consisting in combinig conditionally steps of simulation is used to obtain a non-probabilistic simulation of $U$.
\end{itemize}

\begin{figure}[h]
\centerline{\includegraphics[scale=0.6]{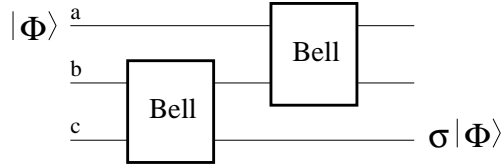}}
\caption{Bell measurement-based teleportation}
\label{fig:telep}
\end{figure}

\begin{figure}[h]
\centerline{\includegraphics[scale=0.6]{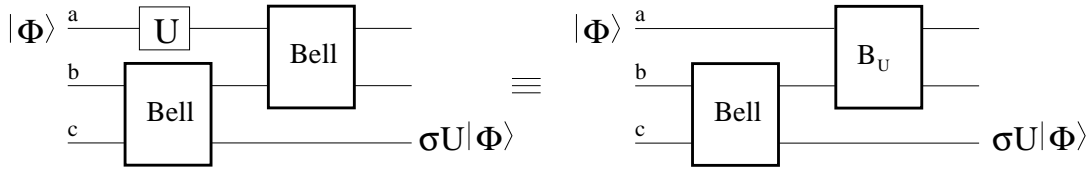}}
\caption{Simulation of $U$ up to a Pauli operator}
\label{fig:gtelep}
\end{figure}

Teleportation can be realized by two successive Bell measurements (figure \ref{fig:telep}), where a Bell measurement is a projective measurement in the basis of the Bell states $\{ \ket{B_{ij}}\}_{i,j\in \{0,1\}}$, where $\ket {B_{ij}} = \frac{1}{\sqrt 2} (\sigma_{z}^i \otimes \sigma_{x}^j) (\ket{00} +\ket {11})$.
A step of simulation of $U$ is obtained by replacing the second measurement by a measurement in the  basis $\{(U^\dagger \otimes Id) \ket {B_{ij}}\}_{i,j\in \{0,1\}}$ (figure \ref{fig:gtelep}).

\begin{figure}[h]
\centerline{\includegraphics[scale=0.5]{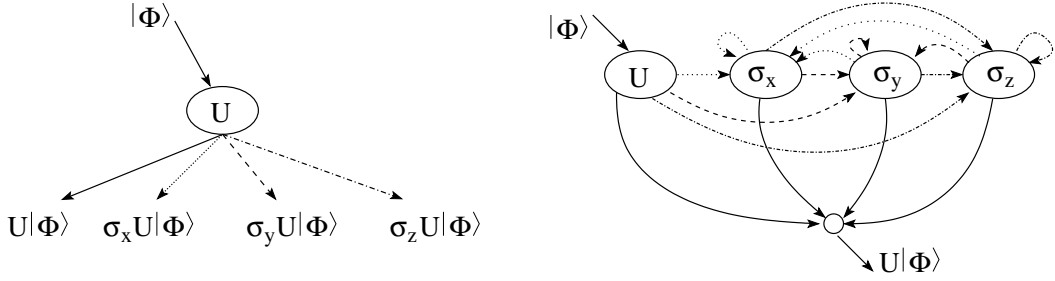}}
\caption{Left: step of simulation abstracted into a probabilistic black box representation -- Rigth: conditional composition of steps of simulation.}
\label{fig:blackbox}
\end{figure}

If a step of simulation is represented as a probabilistic black box (figure \ref{fig:blackbox}, left), there exists a  corrective strategy (figure \ref{fig:blackbox}, {right}) 
which leads to a full simulation of $U$. This strategy consists in  conditionally composing steps of simulation of $U$, but also of each Pauli operator.
A similar step of simulation and strategy are given for the two-qubit unitary transformation $\Lambda X$ (\emph{Controlled}-$X$) in \cite{N01}. Notice that this simulation uses four ancillary qubits.

As a consequence, since any unitary transformation can be decomposed into $\Lambda X$ and one-qubit unitary transformations, any unitary transformation can be simulated by means of projective measurements only.
 More precisely, for any circuit $C$ of size $n$ -- with basis $\Lambda X$ and all one-qubit unitary transformations -- and for any $\epsilon >0$, $O(n\log(n/\epsilon))$ projective measurements are enough to simulate $C$ with probability greater than $1-\epsilon$.

Approximative universality, based on a finite family of projective measurements, can also be considered.  Leung \cite{Leu04} has shown that a family composed of five observables $\mathcal F_{0}=\{Z,X\otimes X, Z\otimes Z, X\otimes Z, \frac{1}{\sqrt 2} (X-Y)\otimes X\}$ is approximatively universal, using four ancillary qubits. It means that for any unitary transformation $U$, any $\epsilon >0$ and any $\delta> 0$, there exists a conditional composition of projective measurements from $\mathcal F_{0}$ leading to the simulation of a unitary transformation $\tilde U$ with probability greater than $1-\epsilon $ and such that $|| U-\tilde U || <\delta$.

 \begin{figure}[h]
\centerline{\includegraphics[scale=0.6]{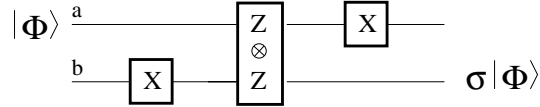}}
\caption{State transfer}
\label{fig:statetrans}
\end{figure}

 \begin{figure}[h]
\centerline{\includegraphics[scale=0.6]{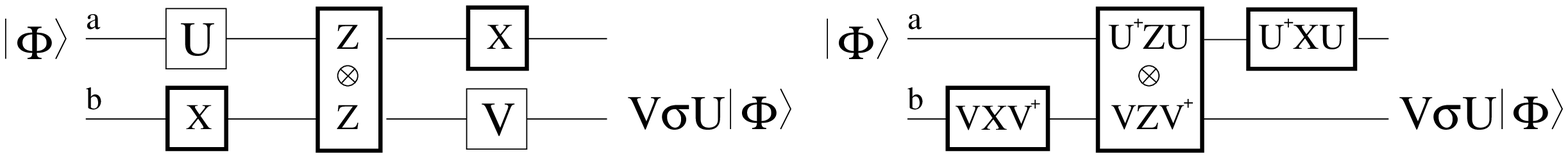}}
\caption{Step of simulation based on state transfer}
\label{fig:statetransgene}
\end{figure}

\begin{figure}[h]
\centerline{\includegraphics[scale=0.6]{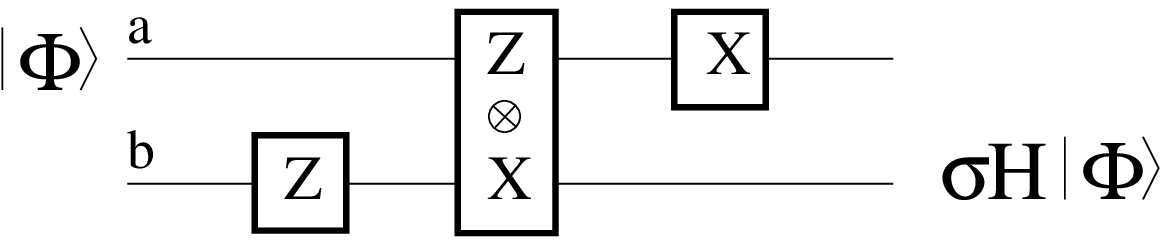}~~~\includegraphics[scale=0.6]{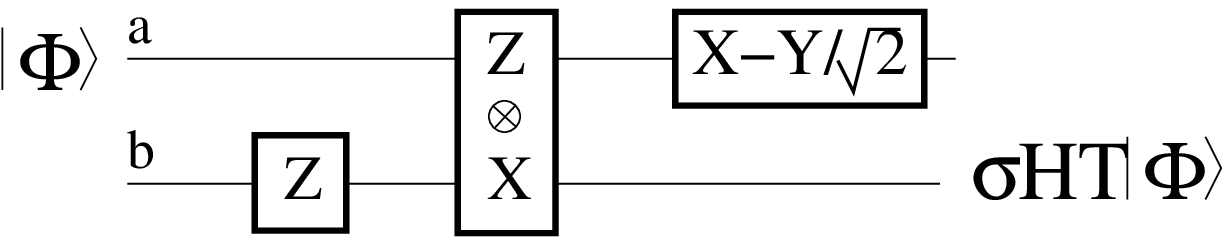}}
\caption{Simulation of $H$ and $HT$ up to a Pauli operator.}
\label{fig:simulH}
\end{figure}

\begin{figure}[h]
\centerline{\includegraphics[scale=0.6]{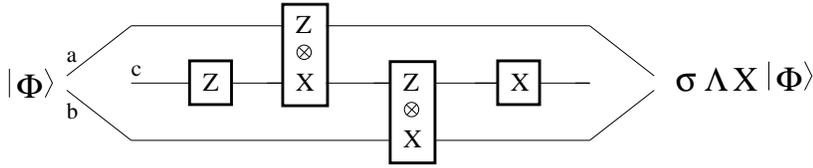}}
\caption{Simulation of $\Lambda X$ up to a Pauli operator.}
\label{fig:simulcnot}
\end{figure}

 In order to decrease the number of two-qubit measurements -- four in $\mathcal F_{0}$ -- and the number of ancillary, an new scheme called \emph{state transfer} has been introduced \cite{Per05a}. The state transfer (figure \ref{fig:statetrans}) replaces the teleportation scheme for realizing a step of simulation. Composed of one two-qubit measurements, two one-qubit measurements, and using only one ancillary qubit, the state transfer can be used to simulate any one-qubit unitary transformation up to a Pauli operator (figure \ref{fig:statetransgene}). For instance, simulations of $H$ and $HT$ -- see section \ref{sec:unituniv} for definitions of $H$ and $T$ -- are represented in figure \ref{fig:simulH}.  Moreover a scheme composed of two two-qubit measurements, two one-qubit measurements, and using only one ancillary qubit can be used to simulated $\Lambda X$ up to a Pauli operator (figure \ref{fig:simulcnot}). Since $\{H,T, \Lambda X\}$ is a universal family of unitary transformations, the family $\mathcal F_{1}=\{Z\otimes X,X,Z, \frac 1{\sqrt 2}(X-Y)\}$ of observables  is approximatively universal, using one ancillary qubit \cite{Per05a}. This result improves the result by Leung, since only one two-qubit measurement and one ancillary qubit are used, instead of four two-qubit measurements and four ancillary qubits. Moreover, one can prove that at least one two-qubit measurement and one ancillary qubit are required for approximative universality. Thus, it turns out that  the upper bound on the minimal resources for measurement-based QC differs form the lower bound, on the number of one-qubit measurements only.

\begin{figure}[h]
\centerline{\includegraphics[scale=0.6]{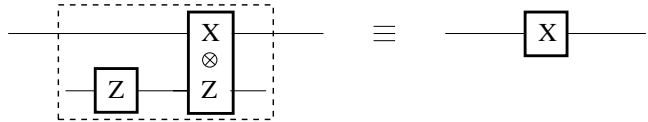}}
\caption{$X$-measurement simulation}
\label{fig:xsimul}
\end{figure}

In \cite{Per04a}, it has been shown that the number of these one-qubit measurements can be decreased, since the family $F_{2} = \{Z\otimes X, Z,  \frac 1{\sqrt 2}(X-Y)\}$, composed of one  two-qubit and only two one-qubit measurements, is also approximatively universal, using \emph{two} ancillary qubit. The proof is based on the simulation of $X$-measurements by means of $Z$ and $Z\otimes X$ measurements (figure \ref{fig:xsimul}). This result leads to a possible \emph{trade-off} between the number of one-qubit measurements and the number of ancillary qubits required for approximative universality.

In this paper, we meanly prove that the family $\mathcal F_{2}$ is approximatively universal, using only one ancillary qubit. Thus, the upper bound on the minimal resources required for approximative universality is improved, and moreover we answer the open question of the trade-off between observables and ancillary qubits. Notice that we prove that the trade-off conjectured in \cite{Per04a} does not exist, but another trade-off between observables and ancillary qubits may exist since the bounds on the minimal resources for measurement-based quantum computation are not tight.

\section{Universal family of unitary transformations}\label{sec:unituniv}

There exist several universal families of unitary transformations, for instance $\{H, T, \Lambda X\}$ is one of them:

\begin{center}
$H=\frac{1}{\sqrt{2}}\left(\begin{array}{cc}
  1 & 1\\
  1 & -1\\
\end{array} \right)$, 
$T=\left(\begin{array}{cc}
  1 & 0\\
  0&e^{\frac{i\pi}{4}}\\
\end{array} \right)$, 
$\Lambda X=\left(\begin{array}{cccc}
  1 & 0&0&0\\
0 & 1&0&0\\
0 & 0&0&1\\
0& 0&1&0
\end{array} \right)$
$\Lambda Z=\left(\begin{array}{cccc}
  1 & 0&0&0\\
0 & 1&0&0\\
0 & 0&1&0\\
0& 0&0&-1
\end{array} \right)$
\end{center}

We prove that the family $\{HT,\sigma_{y}, \Lambda Z\}$ is also approximatively universal:

\begin{theo}\label{thm:dd}
$\mathcal U =\{HT,\sigma_{y}, \Lambda Z\}$ is approximatively universal.
\end{theo}

The proof is based on the following properties. 
Let $R_{\bf n}(\alpha)$ be the rotation of the Bloch sphere about the axis ${\bf n}$ through an angle $\alpha$.
 
 \begin{prop}\label{univer:rotdef}
 If $\;{\bf n}=(a,b,c)$ is a real unit vector, then for any $\alpha$, $R_{\bf n}(\alpha) = \cos (\alpha/2) I -i\sin(\alpha/2)(a\sigma_x+b\sigma_y+c\sigma_z)$.
 \end{prop}

 \begin{prop}\label{univer:approxrot}
 For a given vector ${\bf n}$ of the Bloch sphere, if $\theta$ is an irrational multiple of $\pi$, then for any $\alpha$ and any $\epsilon>0$, there exists $k$ such that $$||R_{\bf n}(\alpha)-R_{\bf n}(\theta)^k)||<\epsilon/3$$
 
 \end{prop}
 
  \begin{prop}\label{univer:decomprot}
 If  $\;{\bf n}$ and ${\bf m}$ are non parallel vectors of the Bloch sphere, then for any one-qubit unitary transformation $U$, there exists $\alpha, \beta, \gamma, \delta$ such that: $$U=e^{i\alpha}R_{\bf n}(\beta)R_{\bf m}(\gamma)R_{\bf n}(\delta)$$
 \end{prop}

 \begin{prop}[W\l odarski \cite{wlod}]\label{univer:irra}
 If $\alpha$ is not an integer multiple of $\pi/4$ and $\cos \beta = \cos^2 \alpha$,
then either $\alpha$ or $\beta$ is an irrational multiple of $\pi$.
 \end{prop}

 \noindent \emph{Proof of theorem \ref{thm:dd}:}
 
 First we prove that any $1$-qubit unitary transformation can be approximated by $HT$ and $\sigma_yHT$.
 Consider the unitary transformations $U_1=T$, $U_2=HTH$, $U_3=\sigma_yHTH\sigma_y$. Notice that $T$ is, up to an unimportant global phase, a rotation by $\pi/4$ radians around $z$ axis on the Block sphere:
 
 \begin{center}
 \begin{tabular}{lclcl}
 $U_1$&$=$&$T$ &$=$&$ e^{-i\pi/8}(\cos(\pi/8)I-i\sin(\pi/8) \sigma_z)$\\
 
 $U_2$&$ =$&$ HTH $&$=$&$   e^{-i\pi/8}(\cos(\pi/8)I-i\sin(\pi/8) \sigma_x)$\\
 
 $U_3$&$ =$&$ \sigma_yHTH\sigma_y $&$= $&$  e^{-i\pi/8}(\cos(\pi/8)I+i\sin(\pi/8) \sigma_x)$
 \end{tabular}
 \end{center}

 Composing $U_1$ and $U_2$ gives, up to a global phase:
 
 \begin{center}
 \begin{tabular}{lcl}
 $U_2U_1$&$=$&$(\cos(\pi/8)I-i\sin(\pi/8) \sigma_x) (\cos(\pi/8)I-i\sin(\pi/8) \sigma_z)$\\
 &$=$&$\cos^2(\pi/8)I -i[\cos(\pi/8)(\sigma_x+\sigma_z)-\sin(\pi/8)\sigma_y]\sin(\pi/8)$
\end{tabular}
 \end{center}
 
According to proposition \ref{univer:rotdef}, $U_2U_1$ is a rotation of the Bloch sphere about an axis along ${\bf n}=(\cos(\pi/8),$ $-\sin(\pi/8),\cos(\pi/8))$ and through an angle $\theta$  defined as a solution of $\cos(\theta/2)=\cos^2(\pi/8)$. Since $\pi/8$ is not an integer multiple of $\pi/4$ but a rational multiple of $\pi$, according to proposition \ref{univer:irra}, a such $\theta$ is an irrational multiple of $\pi$. This irrationality implies that for any angle $\alpha$, the rotation around ${\bf n}$ about angle $\alpha$ can be approximated to arbitrary accuracy by  repeating rotations  around ${\bf n}$ about angle $\theta$ (see proposition \ref{univer:decomprot}). For any $\alpha$ and any $\epsilon>0$, there exists $k$ such that 
 $$||R_{\bf n}(\alpha)-R_{\bf n}(\theta)^k)||<\epsilon/3$$

 Moreover, composing $U_1$ and $U_3$ gives, up to a global phase:
 
 \begin{center}
 \begin{tabular}{lcl}
 $U_3U_1$&$=$&$(\cos(\pi/8)I+i\sin(\pi/8) \sigma_x) (\cos(\pi/8)I-i\sin(\pi/8) \sigma_z)$\\
 &$=$&$\cos^2(\pi/8)I -i[\cos(\pi/8)(-\sigma_x+\sigma_z)+\sin(\pi/8)\sigma_y]\sin(\pi/8)$
\end{tabular}
 \end{center}
 
 $U_3U_1$ is a rotation of the Bloch sphere about an axis along ${\bf m}=(-\cos(\pi/8),\sin(\pi/8),\cos(\pi/8))$ and through the angle $\theta$. Thus, 
 for any $\alpha$ and any $\epsilon>0$, there exists $k$ such that 
 $$||R_{\bf m}(\alpha)-R_{\bf m}(\theta)^k)||<\epsilon/3$$

Since ${\bf n}$ and ${\bf m}$ are non-parallel, any one-qubit unitary transformation $U$, according to proposition \ref{univer:approxrot}, can be decomposed into rotations around ${\bf n}$ and ${\bf m}$ : There exist $\alpha, \beta, \gamma, \delta$ such that $$U= e^{i\alpha}R_{\bf n}(\beta)R_{\bf m}(\gamma)R_{\bf n}(\delta)$$

Finally, for any $U$ and $\epsilon>0$, there exist $k_1,k_2,k_3$ such that 
$$||U - R_{\bf n}(\theta)^{k_1}R_{\bf m}(\theta)^{k_2}R_{\bf n}(\theta)^{k_3}||<\epsilon$$

Thus, any one-qubit unitary transformation can be approximated by means of $U_2U_1$, and $U_3U_1$. Since $U_2U_1=(HT)(HT)$ and $U_3U_1= \sigma_yHTH\sigma_yT= -(\sigma_yHT)(\sigma_yHT)$, the family $\{ HT, \sigma_y\}$ approximates any one-qubit unitary transformation. 

With the additional $\Lambda Z$ gate, the family $\mathcal{U}$ is approximatively universal. $\hfill \Box$

 \section{Universal family of projective measurements}
 
In \cite{Per04a}, the family of observables $\mathcal F_2=\{Z\otimes X, Z, \frac{X-Y}{\sqrt 2}\}$ is proved to be approximatively universal using two ancillary qubits. We prove that this family requires only one ancillary qubit to be universal:

\begin{theo}\label{thm:obsmin}
 $\mathcal F_2=\{Z\otimes X, Z, \frac{X-Y}{\sqrt 2}\}$ is approximatively universal, using one ancillary qubit only. 
  \end{theo}

 The proof consists in simulating the unitary transformations of the universal family $\mathcal U$.
 First, one can notice that $HT$ can be simulated up to a Pauli operator, using measurements of $\mathcal F_{2}$, as it is depicted in figure \ref{fig:simulH}. So, the universality is reduced to the ability to simulate $\Lambda Z$ and the Pauli operators -- Pauli operators are needed to simulated $\sigma_{y}\in \mathcal F$, but also to perform the corrections required by the corrective strategy (figure \ref{fig:blackbox}).

\begin{figure}[h]
\centerline{\includegraphics[scale=0.6]{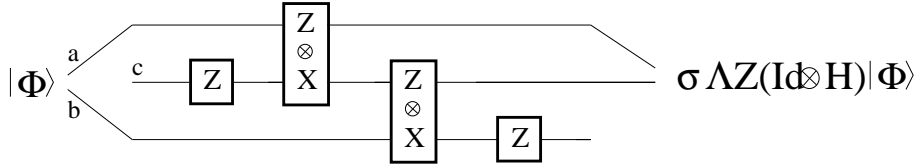}}
\caption{Simulation of $\Lambda Z (Id\otimes H)$}
\label{fig:simulcz}
\end{figure}

  \begin{lemme} \label{lem:simulcz}
  For a given 2-qubit register $a,b$ and one ancillary qubit $c$, the sequence of measurements according to $Z_{c}$, $Z_{a}\otimes X_{c}$, $Z_{c}\otimes X_{b}$, and $Z_{b}$ (see figure \ref{fig:simulcz}) simulates $\Lambda Z (Id\otimes H)$ on qubits $a,b$, up to a Pauli operator. The resulting state is located on qubits $a$ and $c$.
  \end{lemme}

\noindent \emph{Proof:}
One can show that, if the state of the register $a,b$ is $\ket \Phi$ before the sequence of measurements, the state of the register $a,b$ after the measurements is $\sigma \Lambda Z (Id\otimes H) \ket \Phi$, where $\sigma = \sigma_{z}^{s_{1}}\otimes \sigma_{x}^{s_{3}}\sigma_{z}^{s_{2}+s_{4}}$ and $s_{i}$'s are the classical outcomes of the sequence of measurements. \hfill $\Box$

In order to simulate Pauli operators, a new scheme, different from the state transfer, is introduced.

\begin{figure}[h]
\centerline{\includegraphics[width=0.5\textwidth]{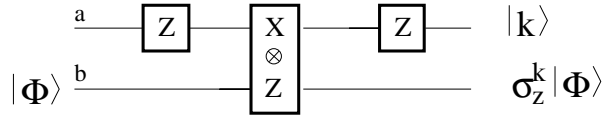}}
\caption{Simulation of $\sigma_z$}
\label{fig:simulZ}
\end{figure}

\begin{lemme}\label{lem:simulZ}
For a given qubit $b$ and one ancillary qubit $a$, the sequence of measurements $Z_{a}$, $X_{a}\otimes Z_{b}$, and $Z_{a}$ (figure \ref{fig:simulZ}) simulates, on qubit $b$, the application of $\sigma_{z}$ with probability $1/2$ and $Id$ with probability $1/2$.

\end{lemme}

 \noindent\emph{Proof:}
 Let $\ket \Phi = \alpha \ket 0 + \beta \ket 1$ be the state of qubit $b$. After the first measurement, the state of the register $a,b$ is $ \ket {\psi_1}=  (\sigma_x^{s_{1}} \otimes Id)\ket 0\otimes\ket\Phi$ where $s_{
 1}\in \{0,1\}$ is the classical outcome of the measurement.
 
 Since $\bra {\psi_1} X\otimes Z \ket{\psi_1} = 0$, the state of the register $a,b$ is:

$$\begin{array}{rcl}
 \ket {\psi_2}&=& \frac{\sqrt 2}2 (\sigma_x^{s_{1}} \otimes Id)(Id +(-1)^{s_{2}}X\otimes Z)\ket 0\otimes\ket\Phi \\
 &=& \frac{\sqrt 2}2 (\sigma_x^{s_{1}}\sigma_{z}^{s_{2}} \otimes Id)(\ket 0\otimes\ket\Phi + \ket 1\otimes (\sigma_z\ket \Phi) \\
 \end{array}$$
 
 Let $s_{3}\in \{0,1\}$ be the outcome of the last measurement, on qubit $a$. If $s_{1}=s_{3}$ then  state of the qubit $b$ is $\ket \Phi$, and $\sigma_{z}\ket \Phi$ otherwise. One can prove that these two possibilities occur with equal probabilities. \hfill $\Box$

\begin{figure}[h]
\centerline{\includegraphics[width=0.5\textwidth]{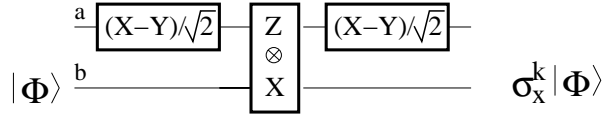}}
\caption{Simulation of $\sigma_x$}
\label{fig:simulX}
\end{figure}

\begin{lemme}\label{lem:simulX}
For a given qubit $b$ and one ancillary qubit $a$, the sequence of measurements $\left(\frac{X-Y}{\sqrt 2}\right)_a$, $Z_{a}\otimes X_{b}$, and $\left(\frac{X-Y}{\sqrt 2}\right)_a$ (figure \ref{fig:simulX}) simulates, on qubit $b$, the application of $\sigma_{x}$ with probability $1/2$ and $Id$ with probability $1/2$.
\end{lemme}

\noindent The proof of lemma \ref{lem:simulX} is similar to the proof of lemma \ref{lem:simulZ}.

\noindent\emph{Proof of theorem \ref{thm:obsmin}:}

First notice that the family of unitary transformations $\mathcal U'=\{HT, \sigma_y, \Lambda Z(I\otimes H)\}$ is approximatively universal since  $\mathcal U=\{HT, \sigma_y, \Lambda Z\}$ is universal.

$HT$ and $ \Lambda Z(I\otimes H)$ can be simulated up to a Pauli operator (lemmas \ref{lem:simulcz}). The universality of the family of observables $\mathcal F_2=\{Z\otimes X, Z, \frac{X-Y}{\sqrt 2}\}$ is reduced to the ability to simulate any Pauli operators. Lemma \ref{lem:simulX} (resp. lemma \ref{lem:simulZ}), shows that $\sigma_{x}$ ($\sigma_{z}$) can be simulated with probability $1/2$, moreover if the simulation fails, the resulting state is same as the original one. Thus, this simulation can be repeated until a full simulation of $\sigma_{x}$ ($\sigma_{z}$). Finally, $\sigma_{y} = i\sigma_z\sigma_{x}$ can be simulated, up to a global phase, by combining simulations of $\sigma_x$ and $\sigma_{z}$. Thus, $\mathcal F_2=\{Z\otimes X, Z, \frac{X-Y}{\sqrt 2}\}$ is approximatively universal using only one ancillary qubit. \hfill $\Box$

\section{Conclusion}

We have proved a new upper bound on the minimal resources required for measurement-based QC: one two-qubit, and two one-qubit observables are universal, using one ancillary qubit only. This new upper bound has experimental applications, but allows also to prove that the trade-off between observables and ancillary qubits, conjectured in \cite{Per04a}, does not exist. This new upper bound is not tight since the lower bound on the minimal resources for this model is one two-qubit observable and one ancillary qubit.

\maketitle
\end{document}